\documentclass[aps,twocolumn,superscriptaddress,longbibliography,nofootinbib]{revtex4-2}

\usepackage{amsmath}
\usepackage{amssymb}
\usepackage{amstext}
\usepackage{amsopn}
\usepackage{amsfonts}
\usepackage{amsxtra}
\usepackage[english]{babel}
\usepackage{graphicx}
\usepackage{bm}
\usepackage{multirow}
\usepackage{dcolumn}
\usepackage{color}
\usepackage{verbatim}
\def\AVS{AV$_{3}$Sb$_{5}$}

\def \kag{kagom{\'e}}
\def \Kag{Kagom{\'e}}

\def\mathbi#1{\ensuremath{\textbf{\em #1}}}
\def\Q{\mathbi{Q}}
\def\QCDW{\ensuremath{\mathbf{Q}_{\text{CDW}}}}
\def\TCDW{\ensuremath{T_{\text{CDW}}}}

\begin{document}

\title{Geometry of the charge density wave in \kag{} metal \AVS{}}

\author{H. Miao} \email{miaoh@ornl.gov}
\affiliation{Material Science and Technology Division, Oak Ridge National Laboratory, Oak Ridge, Tennessee 37831, USA}
\author{H. X. Li}
\affiliation{Material Science and Technology Division, Oak Ridge National Laboratory, Oak Ridge, Tennessee 37831, USA}
\author{W. R. Meier}
\affiliation{Material Science and Technology Division, Oak Ridge National Laboratory, Oak Ridge, Tennessee 37831, USA}
\author{H. N. Lee}
\affiliation{Material Science and Technology Division, Oak Ridge National Laboratory, Oak Ridge, Tennessee 37831, USA}
\author{A. Said}
\affiliation{Advanced Photon Source, Argonne National Laboratory, Argonne, Illinois 60439, USA}
\author{H. C. Lei}
\affiliation{Department of Physics and Beijing Key Laboratory of Opto-Electronic Functional Materials and Micro-devices, Renmin University of China, Beijing, China}
\author{B. R. Ortiz}
\affiliation{Materials Department and California Nanosystems Institute, University of California Santa Barbara, Santa Barbara, California 93106, USA}
\author{S. D. Wilson}
\affiliation{Materials Department and California Nanosystems Institute, University of California Santa Barbara, Santa Barbara, California 93106, USA}
\author{J. X. Yin}
\affiliation{Laboratory for Topological Quantum Matter and Advanced Spectroscopy (B7), Department of Physics, Princeton, New Jersey 08544, USA}
\author{M. Z. Hasan}
\affiliation{Laboratory for Topological Quantum Matter and Advanced Spectroscopy (B7), Department of Physics, Princeton, New Jersey 08544, USA}
\author{Ziqiang Wang}
\affiliation{Department of Physics, Boston College, Chestnut Hill, Massachusetts 02467, USA}
\author{Hengxin Tan}
\affiliation{Department of Condensed Matter Physics, Weizmann Institute of Science, Rehovot 7610001, Israel}
\author{Binghai Yan}
\affiliation{Department of Condensed Matter Physics, Weizmann Institute of Science, Rehovot 7610001, Israel}

\date{\today}

\begin{abstract}

\Kag{} lattice is a fertile platform for topological and intertwined electronic excitations. Recently, experimental evidence of an unconventional charge density wave (CDW) is observed in a Z2 \kag{} metal \AVS{} (A= K, Cs, Rb). This observation triggers wide interests on the interplay between frustrated crystal structure and Fermi surface instabilities. Here we analyze the lattice effect and its impact on CDW in \AVS{}. Based on published experimental data, we show that the CDW induced structural distortions is consistent with the theoretically predicted inverse star-of-David pattern, which preserves the $D_{6h}$ symmetry in the \kag{} plane but breaks the sixfold rotational symmetry of the crystal due to the phase shift between \kag{} layers. The coupling between the lattice and electronic degrees of freedom yields a weak first order structural transition without continuous change of lattice dynamics. Our result emphasizes the fundamental role of lattice geometry in proper understanding of unconventional electronic orders in \AVS{}.

\end{abstract}

\maketitle

\Kag{} lattice is a corner shared triangle network that contains three sites per unit cell \cite{Syozi1951}. The electronic interference between the three sublattices gives rise flat band, van Hove singularity (saddle point) and Dirac-fermion in its band structure. It has been predicted that, near the van Hove filling, the combination of high density of state, sublattice interference and non-local Coulomb interaction may yield unconventional Fermi surface instabilities, such as the $p$-wave charge and spin density waves, $d$-wave Pomeranchuk instability and $f$-wave superconductivity \cite{Wang2013, Thomale2013, feng2021chiral, Thomale_2021,Lin2021,Park2021, Lin2021}. Recently, a three-dimensional charge density wave (CDW) that possibly intertwines with superconductivity is observed in a \kag{} metal \AVS{} (A= K, Cs, Rb) \cite{Ortiz2019, Ortiz2020,jiang2020discovery,Yin2021,tan2021charge,feng2021chiral,zhao2021cascade,liang2021threedimensional,chen2021roton,yu2021concurrence,chen2021double, Li2021,Dai2021,WangNL2021}. While a three-dimensional 2$\times$2$\times$2 superstructure is experimentally identified \cite{Li2021, liang2021threedimensional, Song2021, Oritz2021}, the nature of the CDW and its interplay with the lattice degree of freedom are under rigorous investigations.

%
\begin{figure}
\includegraphics[width=8.5 cm]{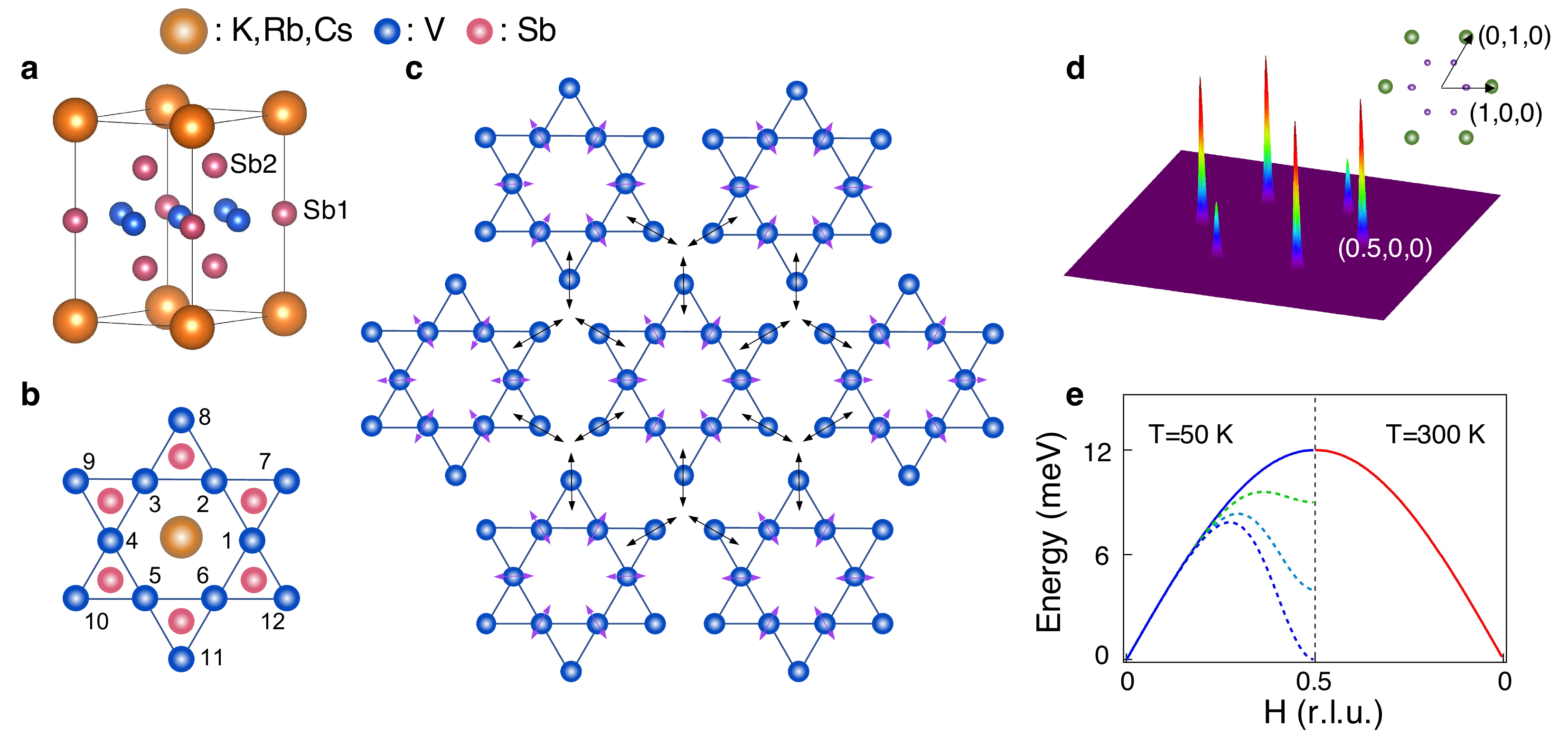}
\caption{Crystal structure and CDW induced lattice distortion of \AVS{}. (a) and (b) side-view and top-view of \AVS{} structure. DFT calculations find that star-of-David (SD) is a local energy minimum at zero temperature, while ISD is the global minimum. (c) shows the SD and ISD lattice distortion of the V-\kag{} lattice. Purple and Black arrows show atomic distortions of V$^{1-6}$ and V$^{7-12}$ (see (b)), respectively. Distortions of V$^{1-6}$ and V$^{7-12}$ are out-of-phase. The ISD pattern corresponds to V$^{1-6}$/V$^{7-12}$ moving toward/outward the center of the V-haxagon. (d) reproduces the STS determined CDW peak intensity near Q=0 \cite{jiang2020discovery}. The inset shows the Bragg (green) and CDW (purple) peaks in the momentum space. The coordinates are shown in reciprocal lattice units. (e) shows the longitudinal acoustic phonon dispersion at 300 and 50 K \cite{Li2021}. The dashed curves show acoustic phonon anomalies that are expected for strong electron-phonon driven CDW \cite{Varma1983,Miao2018,Weber2011,Weber2011_2,Kogar1314}.
\label{Fig1}}
\end{figure}

Figures~\ref{Fig1}a and b show the crystal structure of \AVS{}, which has a space group No.~191 (P6/mmm). The V-Sb slab interlaces with the alkali triangle network along the crystal $c$-axis. Structurally, there are two Sb positions: Sb1 is located at the center of the V-hexagon and Sb2 is sitting above and below the V-triangles. Density functional theory (DFT) calculations found that the the ideal \kag{} structure is energetically unstable and favors an inverse star-of-David (ISD) structure at zero temperature \cite{tan2021charge}. While the ISD distortion of the two-dimensional \kag{} lattice preserves the $D_{6h}$ symmetry, recent scanning tunneling spectroscopy (STS) studies found that the CDW superlattice peaks break the six-fold rotational symmetry (Fig.~\ref{Fig1}d), suggesting a chiral CDW or electronic nematicity \cite{jiang2020discovery,Li2021, Li2021Nematic, Shumiya2021, Xiang2021, chen2021roton}. Magnetoresistence measurements also found evidence of $C_{2}$ symmetry that persists into the superconducting phase \cite{Ni2021,Xiang2021, chen2021roton}. Moreover, as we show in Fig.~\ref{Fig1}e, unlike well-known CDW materials \cite{Rice1975,Varma1983,Hoesch2009,Miao2018,Weber2011,Weber2011_2,Kogar1314,Gruner2018}, CDW in \AVS{} fails to induce acoustic phonon anomalies near the CDW wavevector, \QCDW{}, indicating a strong commensurability effect \cite{Li2021}. Here we explore the CDW by numerically and analytically assessing the structural responses below \TCDW{} in \AVS{}. We show that the three-dimensional ISD structure yields a diffraction pattern that is consistent with x-ray scattering (XRD) and STS measurements \cite{Ortiz2020,Li2021,jiang2020discovery,Li2021, Li2021Nematic, Shumiya2021}. Our analysis supports a CDW in \AVS{} that preserves the $D_{6h}$ in the \kag{}-plane. However, due to the phase shift between \kag{} layers, the CDW breaks the sixfold rotational symmetry, $C_{6}$, and strongly modifies the CDW superlattice peak intensities. Finally, we show that the coupling between CDW and lattice distortion yields a weak first order phase transition \cite{Mu2021,Song2021} that may be responsible for the absence of acoustic phonon anomaly in \AVS{} \cite{Li2021}.

%
\begin{figure}
\includegraphics[width=8 cm]{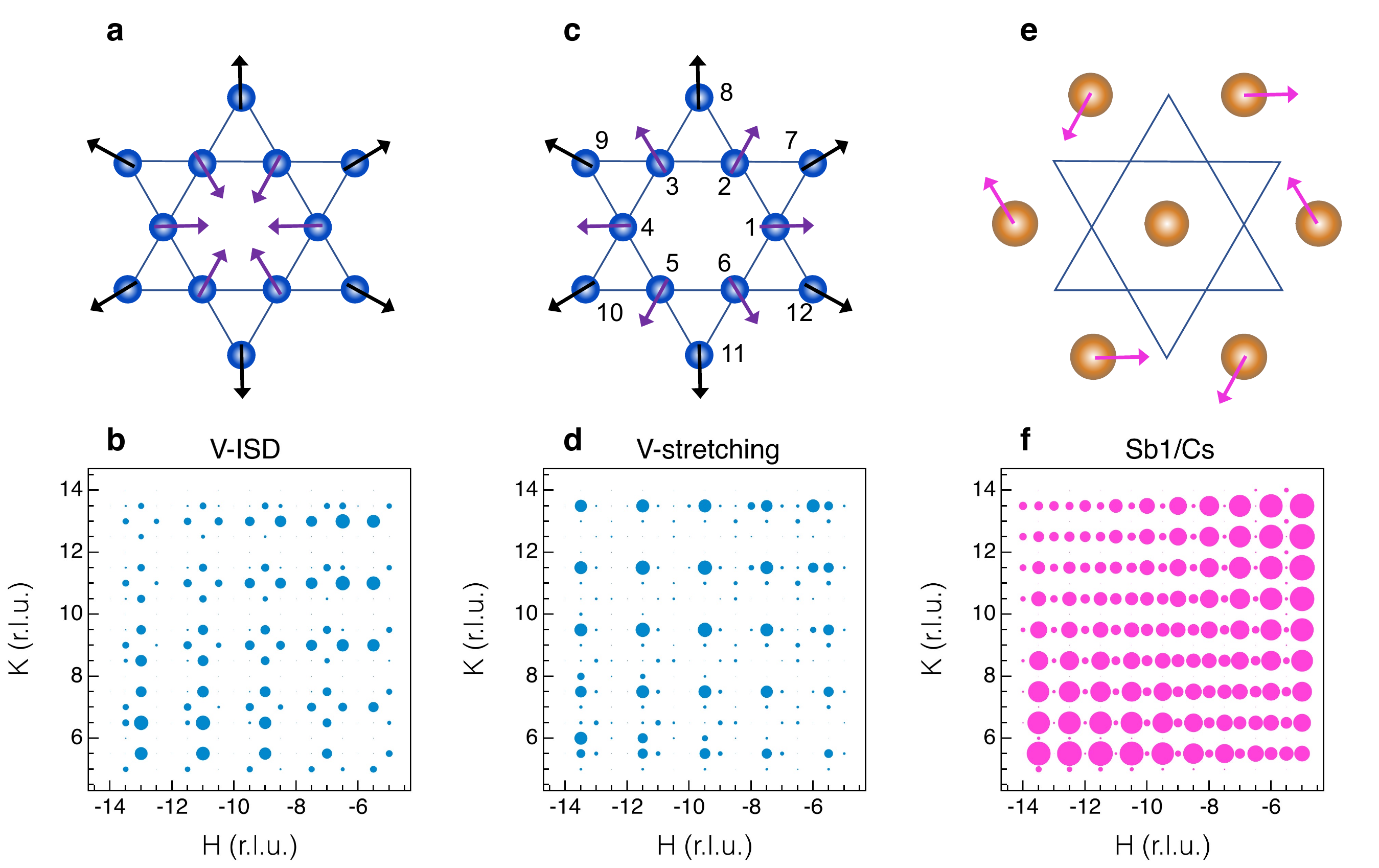}
\caption{CDW induced lattice distortions in real space and their corresponding diffraction patterns in momentum space. (a) and (b) ISD distortion on V-\kag{} sublattice. (c) and (d) V-stretching distortion on V-\kag{} sublattice, both of which preserve the $D_{6h}$ symmetry in the \kag{} plane. (e) and (f) Sb1/Cs distortion that has $C_{3h}$ symmetry. In the diffraction calculations, the lattice distortions were set to be 1\% deviating from the original positions. The size of the dots proportional to the intensity of the diffraction pattern. 
\label{Fig2}}

\end{figure}

The diffraction intensity follows, $I(\mathbi{Q})=|F(\mathbi{Q})|^{2}$, where F(\Q) is the scattering amplitude. For a crystalline material, F(\Q) can be formulated as:

\begin{equation}
\begin{aligned}
F(Q)=\overbrace{\sum_{R}e^{iQ\cdot R_{n}}}^{lattice}\overbrace{\sum_{j}f_{j}(Q)e^{iQ\cdot r_{j}}}^{unit\:cell} =\delta_{Q=G}\overbrace{\sum_{j}f_{j}(Q)e^{iQ\cdot r_{j}}}^{unit\:cell} 
\end{aligned}
\label{Diff}
\end{equation}

\noindent where $\ensuremath{\mathbf{R_{n}}}$ and $\ensuremath{\mathbf{G}}$ are real and reciprocal lattice vectors, respectively. $r_j$ is the $j^{th}$ atomic position in the unit cell. $\ensuremath{\mathbf{Q}}$ is the total momentum transfer and $f_{j}(\ensuremath{\mathbf{Q}})$ is the atomic form factor, which is derived from a Fourier transformation of local density of state (see Appendix A for more details). Below \TCDW{}, the formation of CDW distorts the high-temperature structure and gives rise to superlattice peaks at $\ensuremath{\mathbf{Q}}$=\QCDW{}. 

We first consider the DFT predicted ISD distortions of V-\kag{} lattice. Figure~\ref{Fig2}a schematically shows the ISD distortion, where V$^{1-6}$ and V$^{7-12}$ are breathing out-of-phase with respect to the center of V-hexagon. Figure~\ref{Fig2}b shows the simulated diffraction pattern of ISD shown in Fig.~\ref{Fig2}a. The scattering region is chosen to match previous XRD measurement at $L$=0 plane \cite{Ortiz2020, Li2021}, which captures in-plane atomic distortions. Remarkably, we find that ISD reproduces the key feature of experiment \cite{Ortiz2020, Li2021}, $i.e.$, the CDW peak intensity is significantly larger at $\ensuremath{\mathbf{Q}}$=($Odd$, $Int$+0.5, 0) or ($Int$+0.5, $Odd$, 0), than at $\ensuremath{\mathbf{Q}}$=($Even$, $Int$+0.5, 0) or ($Int$+0.5, $Even$, 0), where $Even/Odd$ and $Int$ represent even/odd integers and integers, respectively. As a comparison, we also calculate two more CDW diffraction patterns: (i) the V-stretching shown in Fig.~\ref{Fig2}c, corresponding to an in-phase breathing of V$^{1-6}$ and V$^{7-12}$ and preserving the $D_{6h}$ symmetry; and (ii) the inversion-symmetry breaking Cs/Sb1 distortion with $C_{3h}$ symmetry as shown Fig.~\ref{Fig1}e. Apparently, these lattice distortions are incompatible with the empirical selection rules and can be excluded for \AVS{}. We shall note that the star-of-David (SD) distortion \cite{tan2021charge,Oritz2021} also captures the main feature of XRD measurement with subtle differences from ISD structure (see Appendix~C). A more detailed x-ray diffraction measurement and structure refinement are required to distinguish these two patterns \cite{tan2021charge}.

\begin{figure}
\includegraphics[width=8.5 cm]{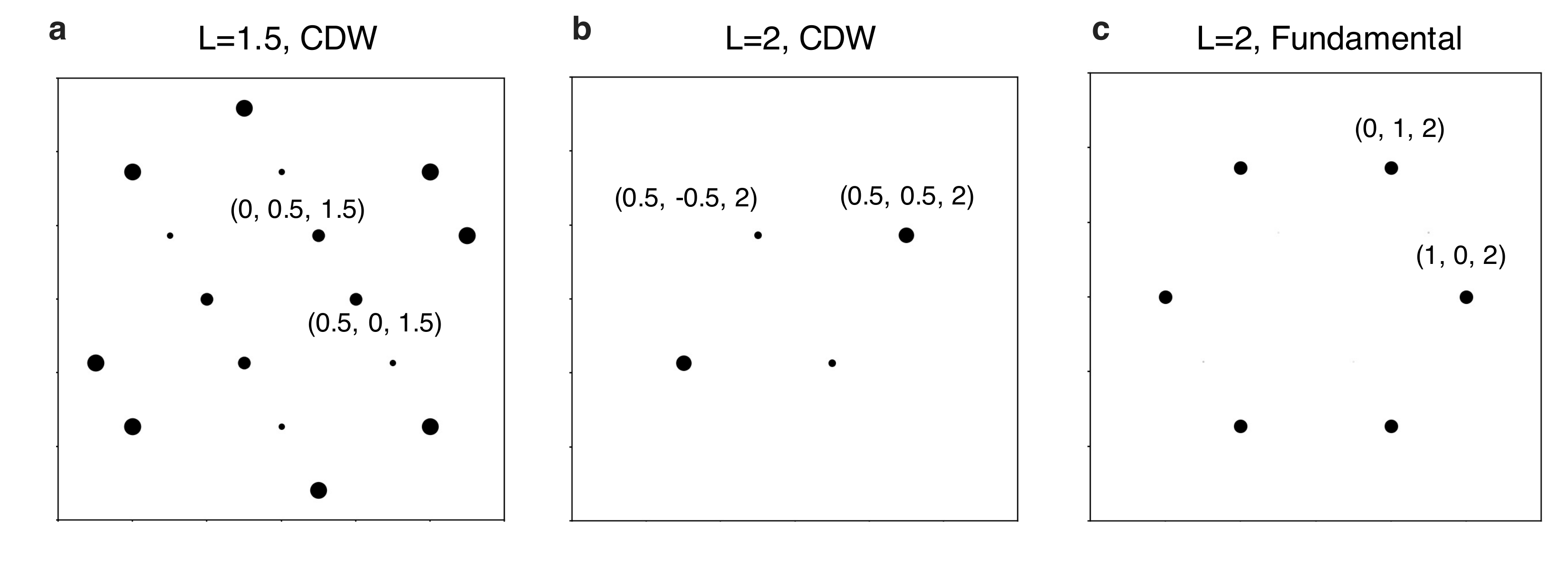}
\caption{$C_{2}$ CDW peak intensity. (a) and (b) show calculated CDW superlattice peak intensity at $L$=1.5 and 2, respectively. (c) shows the fundamental Bragg peak intensity. The calculation is based on the theoretically refined ISD structure \cite{tan2021charge}. The size of the dots proportional to the intensity of the diffraction pattern. 
\label{Fig3}}

\end{figure}

Since the two-dimensional ISD/SD has the $D_{6h}$ symmetry, the CDW superlattice peaks are expected to show $C_{6}$ symmetry. Indeed, based on Eq.~\ref{Diff} and the ISD/SD distortion, we find that:
\begin{equation}
\begin{aligned}
& F(0.5,0,0)=F(0,0.5,0) \\
& =F(-0.5,0.5,0)\approx-2\pi(\delta+\epsilon)
\end{aligned}
\label{CDW_I}
\end{equation}
\noindent where $\delta$ and $\epsilon$ are corresponding to $V^{1-6}$ and $V^{7-12}$ distortions, respectively. Here we assume $|\delta|,|\epsilon|\ll a_{0}=5.4949\AA$. This condition is justified by previous XRD measurement, where the CDW superlattice peaks are 3$\sim$5 orders smaller than their nearby fundamental Bragg peaks \cite{Li2021}. The $C_{6}$ symmetry of the CDW superlattice peak is, however, incompatible with recent STS studies, where only a $C_{2}$ symmetry is observed \cite{jiang2020discovery,Li2021, Li2021Nematic, Shumiya2021}. As we continue to show below, when CDW is three-dimensional (3D), as reported by recent experimental studies \cite{Li2021, Ratcliff2021, yu2021concurrence, Song2021, Oritz2021}, the $C_{6}$ symmetry of the CDW peak intensity naturally breaks down to $C_{2}$. To show the rotational symmetry breaking, we use the theoretically refined 2$\times$2$\times$2 ISD structure of \AVS{} \cite{tan2021charge}. Due to the $\pi$ phase shift between adjacent \kag{} layers, the crystal symmetry is lowered from $D_{6h}$ to $D_{2h}$ \cite{Park2021}. Figure~\ref{Fig3}a and b show the calculated CDW superlattice peak intensity at $L=1.5$ and $L=2$, respectively. In agreement with STS studies \cite{Li2021, Ratcliff2021, yu2021concurrence, Song2021, Oritz2021}, the CDW superlattice peak intensity only shows $C_{2}$ symmetry. In contrast, the fundamental Bragg peak only weakly breaks the $C_{6}$ symmetry due to small lattice distortions. We note that in our simulation, the fundamental Bragg peak intensity is three-orders larger than the CDW peak intensity. 

%

Theoretically, three types of CDW order parameters are predicted for \AVS{}, the onsite CDW, bond CDW and imaginary bond CDW involving flux or loop currents \cite{feng2021chiral, feng2021CDW, Thomale_2021,Ratcliff2021, Park2021}. Due to finite electron-phonon coupling, CDW patterns have to respect the point group symmetry of the lattice. For instance, the Sb/Cs1 distortion shown in Fig.~\ref{Fig2}e is derived from an onsite CDW with $C_{3h}$ symmetry \cite{feng2021CDW}. Our simulations demonstrate that the DFT calculated 2$\times$2$\times$2 superstructure at zero temperature is consistent with experimental observations and therefore support a CDW with $D_{6h}$ symmetry in the \kag{} plane. Furthermore, as we shown in Fig.~\ref{Fig2}a-d, the anti-phase breathing of V$^{1-6}$ and V$^{7-12}$ puts another constraint on the CDW pattern in \AVS{} \cite{feng2021CDW}. Our results, however, do not explain the observed chiral CDW peak intensity in \AVS{} \cite{jiang2020discovery}. Indeed, the chiral CDW keeps the $D_{6h}$ symmetry of the \kag{} plane \cite{feng2021chiral} and hence cannot be distinguished by non-resonant x-ray scattering. Instead, the V $L$-edge resonant x-ray scattering, which selectively enhance electronic excitions from V $3d$-orbital, might be a sensitive probe for this novel electronic order parameter \cite{Achkar2016}. 

\begin{figure}
\includegraphics[width=8 cm]{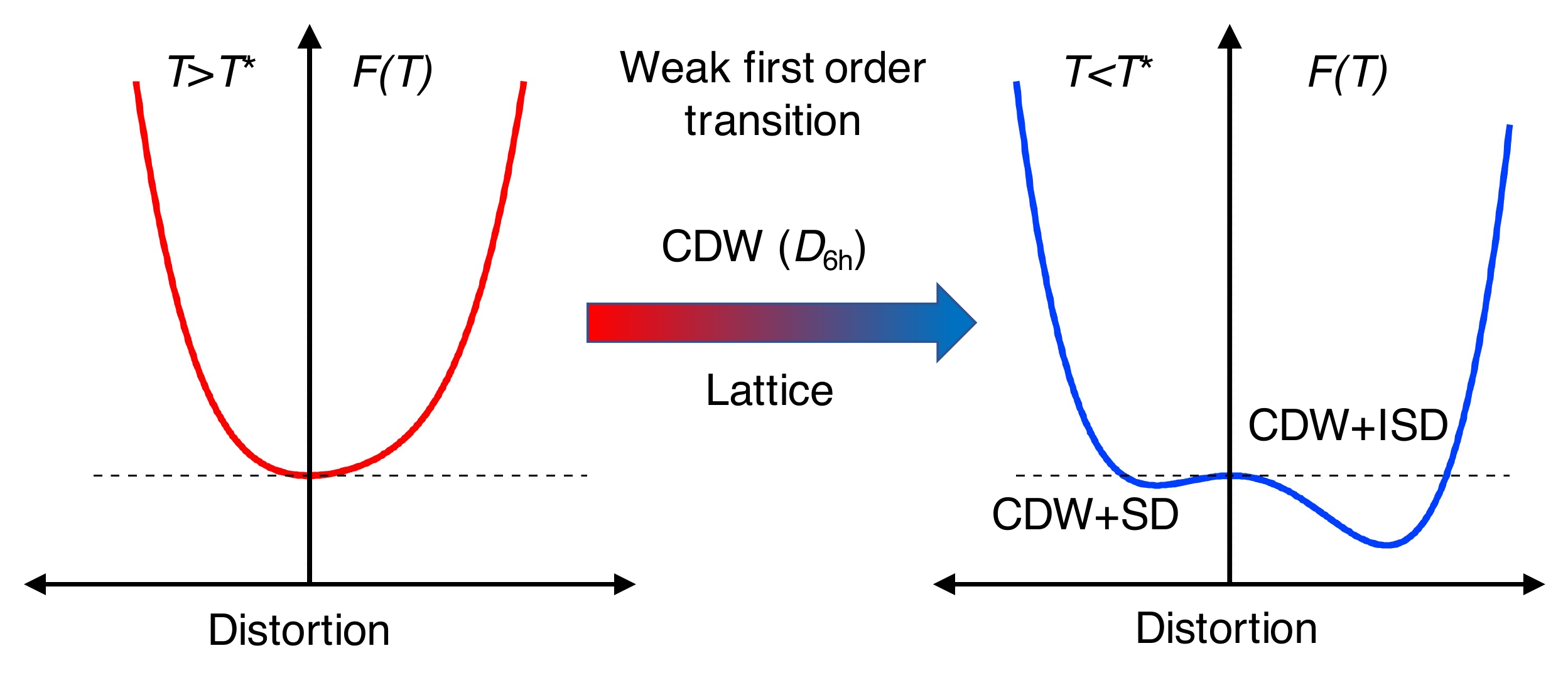}
\caption{CDW-lattice coupling induced weak first order phase transition. At high temperature, the ideal \kag{} lattice is stable and corresponding to the free energy minimum. At zero temperature, ISD is the energy minimum while SD is a local minimum. Near \TCDW{}, the asymmetric lattice free energy adds a cubic term in the CDW free energy through electron-phonon coupling, which, consequently, drive the CDW transition to a weak first order transition.
\label{Fig4}}
\end{figure}

Finally we explore how lattice distortion intertwines with the CDW transition. Previous DFT calculations \cite{tan2021charge} found that the lattice energy is asymmetric with respect to the lattice distortion, $\eta$, at zero temperature (Fig.~\ref{Fig4}). Since CDW always couples with lattice distortions through finite electron-phonon coupling, the asymmetric lattice-distortion energy adds a cubic term in the CDW free energy and leads to a weak first order phase transition. To elaborate it further, we consider an Ising-type Landau free energy on a two-dimensional \kag{} lattice:
\begin{equation}
F(T,\psi)-F_{0}=A(T)\psi^{2}-C\psi^{3}+B\psi^{4}
\label{Free}
\end{equation}

\noindent where $\psi\sim\Delta_{CDW}\sim\eta$. Note the linear term in Eq.~\ref{Free} can be removed by a linear transformation of $\psi$. $\Delta_{CDW}$ is the CDW gap in single particle spectral function. $C$ is a constant that is proportional to the electron-phonon coupling strength. The hysteresis of the first order phase transition is, $\Delta T=\frac{C^{2}}{4AB}$. Experimentally, $\Delta T\sim$1~K \cite{Song2021, Mu2021}, suggesting a weak electron phonon coupling in \AVS{}. Indeed, the calculated electron-phonon coupling constant from ref.~\cite{tan2021charge} is in the weak coupling regime and between 0.3$\sim$0.46 for \AVS{}, supporting our conclusion. Due to the weak first order phase transition, the change of the lattice dynamics is discontinuous near \TCDW{} and possibly intervenes the softening of the CDW phason mode. Together with the strong commensuratbility effect of the 2$\times$2$\times$2 CDW, the phason gap may remain large above \TCDW{} and failed to yield acoustic phonon softening near the CDW wavevector \cite{Li2021}. 

In summary, we explored the intricate interplay between lattice geometry and CDW in \kag{} metal \AVS{}. We prove that the ISD distortion reproduces the XRD and STS measurements. We showed that the coupling between lattice distortions and CDW induces a weak first order transition without continuous phonon softening in \AVS{}.

%

Acknowledgements: We thank Kun Jiang for stimulating discussion on the interplay between CDW patterns and lattice distortions.This research was sponsored by the U.S. Department of Energy, Office of Science, Basic Energy Sciences, Materials Sciences and Engineering Division and by the Laboratory Directed Research and Development Program of Oak Ridge National Laboratory, managed by UT-Battelle, LLC, for the U. S. Department of Energy. H.C.L. was supported by National Natural Science Foundation of China (Grant No. 11822412 and 11774423), Ministry of Science and Technology of China (Grant No. 2018YFE0202600 and 2016YFA0300504) and Beijing Natural Science Foundation (Grant No. Z200005). Z.Q.W is supported by the U.S. Department of Energy, Basic Energy Sciences Grant No. DE-FG02-99ER45747. B.Y. acknowledges the financial support by the Willner Family Leadership Institute for the Weizmann Institute of Science, the Benoziyo Endowment Fund for the Advancement of Science, Ruth and Herman Albert Scholars Program for New Scientists, the European Research Council (ERC) under the European Union’s Horizon 2020 research and innovation programme (Grant No. 815869) and ISF MAFAT Quantum Science and Technology (2074/19). 

\appendix 

\section{Atomic structure factor}
The atomic form factor is a Fourier transform of a spatial density distribution of the scattering object. It is defined as: 
\begin{equation}
f(\mathbi{Q})=\int\rho(\mathbi{r})e^{i\mathbi{Q}\cdot\mathbi{r}}d^{3}\mathbi{r}
\label{FormF}
\end{equation}
where $\rho(\mathbi{r})$ is the real space electron density. For non-resonant x-ray scattering, the atomic form factor is well approximated by a sum of Gaussians of the form:

\begin{equation}
f(\mathbi{Q})=\sum_{i=1}^{4}a_{i}exp(-b_{i}(\frac{\mathbi{Q}}{4\pi})^{2})+c
\label{FormFApp}
\end{equation}
The coefficients in Eq.~\ref{FormFApp} can be found in \cite{Brown2006}. When considering the lattice vibrations, the atomic form factor will be modified to:
\begin{equation}
f^{DW}(\mathbi{Q})=f(\mathbi{Q})e^{-\frac{1}{2}\mathbi{Q}^{2}\langle u_{Q}^{2}\rangle}\equiv f(\mathbi{Q})e^{-M}
\label{FormF}
\end{equation}
where $\langle u_{Q}^{2}\rangle$ is the time averaged mean of squared atomic displacement. In our calculation, Debye-Waller factor has been neglected.

\section{Structural domain}
We consider the 2$\times$2$\times$2 superstructure which involves a $\pi$-phase shift between the ISD distorted \kag{} layers. Assuming the scattering pattern of the two-dimensional ISD structure is $I^{2\times 2}(\mathbi{Q})$, the scattering intensity of 2$\times$2$\times$2 can be written as:

\begin{equation}
I^{2\times 2\times 2}(\mathbi{Q})=I^{2\times 2}(\mathbi{Q})*|(1+e^{i\mathbi{Q}\cdot\mathbi{T}})|^{2}
\label{Domain}
\end{equation}
where $\mathbi{T}_{a,b,c}=$(1, 0, 1) or (0, 1, 1) or (1, -1, 1) in the high temperature reciprocal lattice unit. For a measurement in a single structural domain, such as STS, Eq.~\ref{Domain} post strong selection rule. For instance, take $\mathbi{T}_{a}=$(1, 0, 1), the CDW peak at (0.5, 0, 0) is actually forbidden. For a measurement that covers multiple domains, such x-ray measurement:

%
\begin{figure}
\includegraphics[width=8.5 cm]{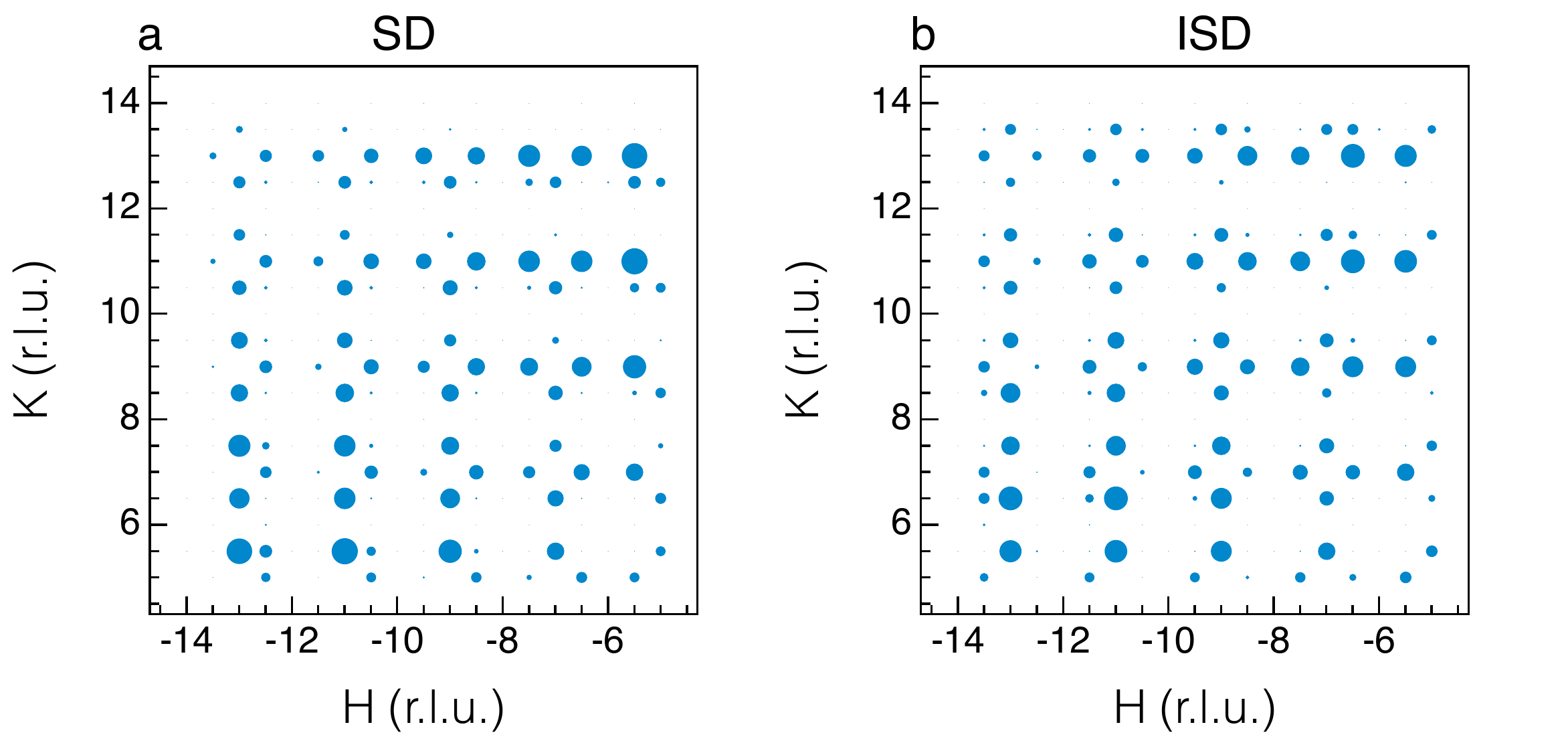}
\caption{Simulated diffraction intensity. (a) V-SD distortion and (b) V-ISD distortion. In the diffraction calculations, the lattice distortions were set to be 1\% deviating from the original positions. The size of the dots proportional to the intensity of the diffraction pattern.
\label{Fig5}}

\end{figure}
%
\begin{equation}
\begin{aligned}
& I^{2\times 2\times 2}(\mathbi{Q}) \\
& =I^{2\times 2}(\mathbi{Q})*(|1+e^{i\mathbi{Q}\cdot\mathbi{T}_{a}}|^{2}+|1+e^{i\mathbi{Q}\cdot\mathbi{T}_{b}}|^{2}+|1+e^{i\mathbi{Q}\cdot\mathbi{T}_{c}}|^{2}) \\
& \propto I^{2\times 2}(\mathbi{Q})
\end{aligned}
\label{Domain}
\end{equation}
Therefore, the CDW superlattice peaks determined by maulti-domain measurements will be similar to the 2$\times$2 CDW.

\section{Star-of-David distortion}

Figure~\ref{Fig5} compares diffraction patterns of SD and ISD distortions. While the both distortions capture the empirical diffraction selection rules, SD and ISD show subtle differences, for instance, the relative intensity between (H, 13.5, 0) and (H, 12.5 0) are opposite for SD and ISD.

\bibliography{ref}
\end{document}